\documentstyle [12pt] {article}
\begin{document}
\title {Geometrization of Some Epidemic Models }
\maketitle
\begin{center}
{\bf {M.E.Kahil{\footnote{Mathematics Department, Modern Sciences and Arts University, Giza, EGYPT\\
e.mail: kahil@aucegypt.edu}}} }
\end{center}
\abstract{
Recently, the behavior of different epidemic models and their relation both to different types of geometries and to some biological models has been revisited . Path equations representing the behavior of epidemic models and their corresponding deviation vectors are examined.  A comparison between paths  and their deviation vectors in Riemannian and Finslerian Geometries is presented.}

\section{Introduction}

In this study, we are going to describe a specific susceptive-infective (SI)-model of epidemics  using Antonelli's approach of describing two species in ecology geometrically [1]. This approach has been successfully used in case of wild rabbit disease as well as coral-starfish equilibrium[2-5].  Also, allometric relationship between net production biomass and the amount of secondary compound plants produce to defend against herbivores can be geometrized as well as corresponding interaction condiments are taken to be constant.
 The aim of this work  is to apply Antonelli's idea of imposing geometrical paths defined in  metric spaces extracted from stochastic allometry space in Riemannian geometry  to obtain a better non-linear regression formula with geometrical origin [2] . The importance of allometry space, due to Antonelli's approach, is to obtain a space having a positive definite metric with negative curvature acting as a tool to examine the behavior of growth curves.  Thus,  any additive terms associated with geometrical structures is useful for adjusting a proper path describing the behavior of any epidemic curve. Since 1990s Antonelli's approach has been extending to include other types of geometries such as Finsler geometry rather than Riemannian[3]. Consequently, Antonelli and Bradbury (1997) have used Gompertz growth and allometric relationships to build a dynamical theory of ecology, evolution and development in colonial organisms using Finslerian Geometry.

Accordingly, geometrization of SI-model can be considered as an introductory step to the geometric version of SIR-model and others. Moreover, this approach has been extended to include Finslerian Geometry due to its richness of imposing many interacting factors geometrically rather than the Riemannian one [5]. The importance of geometrizing such a model is to construct the path equations and their corresponding  path deviation equations to know the behavior of the growth curves and their effect after a slight perturbation on them. This can be seen by studying the stability conditions from examining the motion of path deviation equations. 
In the present work, we are going to utilize the nonlinear version of allometry, Riemannian and Finslerian, of Antonelli  for obtaining the relevant equations of epidemics in  SI-models using different types of geometries with some details on its extension to include SIR models and others by increasing the dimension of the manifold. This type of extending dimensions to examine various epidemic models will be examined in our future work.

\section{Historical Background}
\subsection{ SI-Model of Epidemics}
It is well known that the simplest model of describing population growth is stemmed from the famous prey-predator model cf[7] can be expressed as follows:
$$
\frac{dS}{dt} =-\alpha_{1} SI ,  \eqno{[1.a]}
$$
and
$$
 \frac{dI}{dt} =\alpha_{1} SI,   \eqno{[1.b]}
$$
where$S(t)$ the susceptive class $I(t)$, the infected class and $\alpha_{1}$ s a parameter. Equations (1.a)and(1.b) can be related together using the following condition:
$$
 S(t)+ I(t)=N   \eqno{[2]}
$$
where $N$ stands  for the total number of the population.
The above model has been modified, Lotika-Volterra model , to accommodate some other factors based on some oscillations either internally or externally. In this study, it is sufficient to display the amended version owing to this extra factor to take the following form [8]

$$
\frac{dS}{dt} =-\alpha_{1} SI + \alpha_{2} S \eqno{[3.a]}
$$
and
$$
\frac{dI}{dt} =\alpha_{1} SI + \alpha_{2} I  \eqno{[3.b]}
$$
where $\alpha_{2}$ is another a parameter with constant coefficient, used to measure this interaction.
From this perspective , it is vital to construct similar sets of path  equations  using Riemannian and Finslerian geometries respectively.

\subsection{The Concept of Allometry}

In 1936, Sir Julian Huxley introduced the concept of allometry, or the experimental study of relative growth of parts of animals, via log -log plots of measurements of morphological characteristics of individuals in a Euclidean geometry. These plots resulted straight lines via statistical method of least squares. Around the same time Sir Joseph Needham made some  experimental work to show that straight lines allometries.

In 1944,  J. Kittredge used  the same concept  to estimate the crown biomass of trees in forest stand by measuring the trunk girth, or diameter  at breast height. Again some experiments of botanist J. Harper in 1962 who found Gompertz curves the best ones to fit for describing the growth of simple aquatic plants.

By 1965  Laird studied  on vertebrate growth using  Huxlely's allometric law as well. Several applications of concept of allometry have been discussed in detail. Recently, Karl Niklas (1994) has recorded a variety growing  plants and flowers satisfying the allometric concept as well [6].

\section{Geometerization of Epidemics}

\subsection{Underlying Geometry of SI-model: The Riemmanian Version}
It is well known that, Antonelli and Voorhees (1974) have suggested the following metric in order to define geometrically the behavior of growth curves using allometric space[7].
$$
g_{ij} = e^{-2\alpha_{k}x^{k} }\delta_{ij},  \eqno{[4]}
$$
and its affine connection is given by
$$
\Gamma^{i}_{. jk} = \frac{1}{2}g^{il}( g_{jk,l}+g_{ik,l}-g_{jk,l}), \eqno{[5]}
$$
where $g^{ij}$ is the the matrix inverse of $g_{ij}$. Suppose $\alpha_{i}$ is a constant vector to define the Christoffel symbol in the following way:
$$
\Gamma^{i}_{. ii} = -\alpha_{i} = constant \eqno{[6.a]}
$$
$$
\Gamma^{i}_{. jj} = \alpha_{i} = constant,i \not= j \eqno{[6.b]}
$$
$$
\Gamma^{i}_{. ji} =0 ,  i \not= j   \eqno{[6.c]}
$$

$$
\Gamma^{i}_{. jk} = \Gamma^{i}_{. ij} = \alpha_{i} = constant  i \not= j \not= k \eqno{[6.d]}
$$

Thus, the metricity condition
$$g_{ij;k}=0  \eqno{[7]}$$
becomes
$$
{e^{-2\alpha_{m}x^{m}}\delta_{ij}}_{k} - e^{-2\alpha_{m}x^{m}}\delta_{in}\Gamma^{nj}_{k}- e^{-2\alpha_{m}x^{m}}\delta_{nj}\Gamma^{n}_{ki}=0,
$$
$$
-2\alpha_{k}\delta_{ij} - e^{-2\alpha_{m}x^{m}}\delta_{in}\Gamma^{nj}_{k}- e^{-2\alpha_{m}x^{m}}\delta_{nj}\Gamma^{n}_{ki}=0,
$$
$$
\gamma_{k}\delta_{ij} =\delta_{in}\Gamma^{n}_{kj}+\delta_{jn}\Gamma^{n}_{ik}
$$
We can find out that the Riemann-Christoffel Curvature
$$ R^{c}_{abd} = \Gamma^{c}_{ad,b}-\Gamma^{c}_{ab,d} +\Gamma^{m}_{ad}\Gamma^{c}_{mb}- \Gamma^{m}_{ab}\Gamma^{c}_{md} \eqno{[8.a]} $$ has reduced to

$$ R^{c}_{abd} = \Gamma^{m}_{ad}\Gamma^{c}_{mb}- \Gamma^{m}_{ab}\Gamma^{c}_{md} \eqno{[8.b]}
 $$
Thus the non vanishing components of the curvature tensor become
$$
R^{i}_{jil}=\alpha_{i}\alpha_{l}, i\neq j\neq l \eqno{[9.a]}
$$
 $$
 R^{i}_{jij}=\sum_{k\neq i,j} (\alpha_{k})^2, i\neq j \eqno{[8.b]}
 $$
 $$
 R^{i}_{jjl}=\alpha_{i}\alpha_{j}, i\neq j\neq l. \eqno{[8.c]}
 $$

\subsection{Equations of Epidemics in  Riemannian version of SI-Model }
It is well known that path equations of any growth curves can be obtained from the following the following applying variation principle on  Lagrangian functions. This trend of imposing biological action principle has been stemmed prior to Euler by Mauphritis [2-7],

$$
L=g_{ab}U^{a}U^{b} \eqno{[10]}
$$
 However, in an approach to obtain path and path deviation equations from one single lagrangian using the Bazanski Lagrangian
 [10] which has the advantage that we obtain path and path deviation equations from the same Lagrangian:
$$
L = g_{_{ab}}{U^{a}} {\frac{D \Psi^{b}}{Dt}}\eqno{[11]}
$$
where $a,b=1,2,3,..n$.
By taking the variation with respect to the deviation vector $\Psi^{c}$ and the tangent vector $U^{c}$, and $\frac{D\Psi^{\alpha}}{Ds}$ is covariant derivative with respect to a parameter $s$.  From this perspective, it can easily be found that  $$
\frac{dU^{c}}{dS}+ \Gamma^{c}_{ab} {U ^{a}}{U^{b}}=0\eqno{[12]}
$$
and
$$
\frac{D^2 \Psi^{c}}{Dt^2}= R^{c}_{abd}U^{a}U^{b}\Psi^{d}\eqno{[13]}
$$

The path and path deviation equations of the SI-model can be obtained from the following Bazanski Lagrangian
$$
L= g_{\mu \nu} U^{\mu} \frac{D \Psi^{\nu}}{Dt} \eqno{[14]}
$$

By taking the variation with respect to the deviation vector $\Psi^ {\sigma} $to get the following path equation
$$
\frac{d S}{ dt}+ \Gamma^{1}_{11}{ S}^2 + \Gamma^{1}_{22}{ I}^2 + 2\Gamma^{1}_{12} {S}{ I} =0, \eqno{[15.a]}
$$
and
$$
\frac{d I}{ dt}+ \Gamma^{2}_{11}S^2 + \Gamma^{2}_{22}{ I}^2 + 2\Gamma^{2}_{12} { S}{ I} =0,\eqno{[15.b]}
$$
While  taking the variation with respect to velocity vector $U^{\sigma}$ to get
 their corresponding path deviation equations :
$$
\frac{D^2\Psi_{S}}{ Dt^2} =   R^{1}_{112} { S} { I} \Psi_{I}  + R^{1}_{121}{ S} {I} \Psi_{S}  +R^{1}_{212} { S} { I} \Psi_{I}+ R^{1}_{221}S I \Psi_{S}, \eqno{[16.a]} $$
and
$$
\frac{D^2\Psi_{I}}{ Dt^2} =   R^{2}_{112} { S} { I} \Psi_{I}  + R^{2}_{121}{ S} { I} \Psi_{S}  +R^{2}_{212} { S} { I} \Psi_{I}+ R^{2}_{221}{ S} {I} \Psi_{S}, \eqno{[16.b]}
$$

i.e.

$$
\frac{D^2\Psi_{S}}{ Ds^2}  = (\alpha_{1}^2+ 2 \alpha_{1}\alpha_{2}+ \alpha_{2}^2)   S I ( \Psi_{S}  - \Psi_{I}) \eqno{[16.c]}  ,
$$
and
$$
\frac{D^2\Psi_{I}}{ Ds^2} =  (\alpha_{1}^2+ 2 \alpha_{1}\alpha_{2}+ \alpha_{2}^2)  S I( \Psi_{S}- \Psi_{I})\eqno{[16.d]}
$$

 Consequently, we can suggest  the path path and path deviation equations of a modified SI model can be following Lagrangian:

$$
L= g_{\mu \nu} U^{\mu} \frac{D \Psi^{\nu}}{Dt} +\lambda \Psi^{\mu}U^{\mu}  \eqno{[17]}
$$

to give,
$$
\frac{dU^{a}}{dS}+ \Gamma^{a}_{b}{c} {U ^{b}}{U^{c}}=   \lambda_{a} U^{a}, \eqno{[18.a]}
$$

i.e

$$\frac{d S}{dt}+ { \Gamma}^{1}_{11} S^2 + 2 { \Gamma}^{1}_{12}SI+ { \Gamma}^{1}_{22}I^2= \lambda S \eqno{[18.b]}
$$
and

$$
\frac{dI}{dt}+ { \Gamma}^{2}_{11} S^2 + 2{ \Gamma}^{2}_{12} SI +  \Gamma^{2}_{22}I^2= \lambda I, \eqno{[18.c]}
$$
in which can be expressed as follows

$$
\frac{d S}{dt}+ 2 \beta  S  I -\alpha I^2 + \alpha S^2= \lambda  S \eqno{[18.d]}
$$
and
$$
\frac{d I}{dt}+ 2(\alpha- \beta L) S  I -(\beta + \alpha L) S+ (\alpha- \beta L)I^2= \lambda  I  .\eqno{[18.e]}
$$

And their corresponding path deviation equations become as follows:
$$
\frac{D^2 \Psi^{a}}{Dt^2}= R^{a}_{bcd}U^{b}U^{c}\Psi^{d} + \lambda \frac{D \Psi^{a}}{Dt} \eqno{[19.a]}
$$

$$
\frac{D^2\Psi_{S}}{ Ds^2}+ \lambda \frac{D \Psi_{S}}{Dt} =   R^{1}_{112} { S}  { I} \Psi_{I}  + R^{1}_{121}{  S} {  I} \Psi_{S}  +R^{1}_{212} { S} { I} \Psi_{I}+ R^{1}_{221} { S} {  I} \Psi_{S}  , \eqno{[19.b]}
$$
and
$$
\frac{D^2\Psi_{I}}{ Ds^2}+ \lambda \frac{D \Psi_{I}}{Dt} =   R^{2}_{112} { S } {\dot I} \Psi_{I}  + R^{2}_{121} { S}  {\dot I} \Psi_{S}  +R^{2}_{212} { S}  { I} \Psi_{I}+ R^{2}_{221} { S} { I} \Psi_{S} ,\eqno{[19.c]}
$$
which can be expressed as follows:
$$
\frac{D^2\Psi_{S}}{ Ds^2}+ \lambda \frac{D \Psi_{S}}{Dt} = (\alpha_{1}^2+ 2 \alpha_{1}\alpha_{2}+ \alpha_{2}^2)   S I ( \Psi_{S}  - \Psi_{I})   ,\eqno{[19.d]}
$$
and
$$
\frac{D^2\Psi_{I}}{ Ds^2}+ \lambda \frac{D \Psi_{I}}{Dt} =  (\alpha_{1}^2+ 2 \alpha_{1}\alpha_{2}+ \alpha_{2}^2)  S I( \Psi_{S}- \Psi_{I}) \eqno{[19.e]}
$$

\subsection{Berwald type}
The metric tensor of Berwald type in Finslerian geometry can be described as follows
$$g_{y}=\dot\partial_{i}\dot\partial_{j}(\frac{1}{2}F^2)$$
such a metric tensor is called a Minkowoski space with Finslerian norm $ds=f(dx^1,dx^2)$.
Thus, its corresponding geodesic can be described as
$$
\frac{d^2x^i}{ds^2}+ \hat{\Gamma}^{i}_{jk}  (x,y) =0 \eqno{[20]}
$$
such that
$$
\hat \Gamma^{i}_{jh}= \frac{1}{2}g^{im} ({\delta}_h g_{jm}+\delta_{j}g_{mh}-\delta_{m} g_{jh}), \eqno{[21]}
$$
where  $\delta_{h}$ is a partial derivative with respect to non-linear connection.
$$
\delta_{h} = \partial_{h} - \Gamma^{i}_{jh}(x)\frac{\partial}{\partial y^{h}},\eqno{[22]}
$$
$$
\hat C^{i}_{. jk} = \frac{1}{2}g^{il}( \frac {\partial g_{lk}}{\partial y^{j} }+\frac {\partial g_{il}}{\partial y^{k}}-\frac{\partial g_{jk}}{\partial y^{l}})Ò\eqno{[23]}
$$

The Berwald type has some associated curvatures are defined in the following way [7]:
$$
B^{i}_{kjh}= R^{i}_{kjh} + G^{r}_{j}D^{i}_{rhk}-G^{r}_{k}D^{i}_{rhj} \eqno{[24]}
$$
where
$$
G^{i}_{j} = \hat{\Gamma}^{i}_{jk}y^{j} -\frac{1}{2}g^{im} \hat{\Gamma}^{l}_{jk}\frac{\partial g_{ml}}{\partial y^{l}}y^{j}y^{k} \eqno{[25.a]}
$$
and
$$
D^{i}_{kjh}= \frac{\partial^{3}H^{i}_{2}}{\partial y^{j} \partial y^{k} \partial y^{l}}, \eqno{[25.b]}
$$
such that
$$
\hat {\Gamma}^{i}_{\alpha_{1}\alpha_{2}\alpha_{3}....\alpha_{m} }=\frac{1}{m!} \frac{\partial^{m} H_{(m)}}{\partial y^{\alpha_{1}} \partial y^{\alpha_{2}}  \partial y^{\alpha_{3}}.....\partial y^{\alpha_{m}}} \eqno{[25.c]}
$$

\subsection{ SI-Model in Finsler Geometry}
In 1991 Antonelli  developed  such a metric for 2-dimensional Berwald space with locally constant coefficients to become [3]
$$
F^2 = {e}^{2 \alpha_{i} X^{i}(L^2+1)}((\dot X^1)^2+(\dot X^2)^2), i=1,2 \eqno{[26]}
$$
where $X^1$ and $X^2 $ are Cartesian coordinates on $R^2$ and $L$ acts as a perturbation parameter.
The above relation can be related to Riemananian geometry by relaxing the term $L$
$$
\bar g_{ij}(X, \dot X) = {e}^{2 \alpha_{i} X^{i}(L^2+1)}g_{ij} \eqno{[27]}
$$

The above system of equations together with their corresponding deviation vector equations can be obtained from taking the action principle to the following Lagrangian.
$$
L_{BF}=g_{ij} (x,y) X^{i} \frac{\hat{D} \hat{\Psi}^{j}}{\hat{D} t} \eqno{[28]}
$$
Such that
$$
\frac{\hat D \hat \Psi^i}{\hat D t} = \frac{d \hat{\Psi}^i}{dt}+ \hat \Gamma^{i}_{jh} \hat{\Psi}^{j} U^{h} + C^{i}_{jh} \hat\Psi^{j} V^{h},\eqno{[29]}
$$
where $$V^{h} = \frac{\delta y}{\delta t} .\eqno{[30]}$$
Accordingly, geodesic equation may be as follows
$$
\frac{d^2x^{i}}{dt^2} + \hat \Gamma^{i}_{jh}(x,y) y^{j}y^{h} = 0 ,
$$
i.e.
$$
\frac{d S}{dt}+ 2(\beta+ \alpha L) S  I +(\beta L -\alpha) I^2 + (\alpha- \beta L) S^2= 0 \eqno{[31.a]}
$$
and
$$
\frac{d I}{dt}+ 2(\alpha- \beta L)\dot S \dot I -(\beta + \alpha L)\dot S+ (\alpha- \beta L)I^2= 0 \eqno{[31.b]}
$$
and its deviation equation becomes
$$
\frac{\hat{D}^2 \hat{\Psi}^{a}}{\hat{D}t^2}= B^{a}_{bcd}U^{b}U^{c}\hat{\Psi}^{d}, \eqno{[32.a]}
$$

to become
$$
\frac{\hat D^2\Psi_{S}}{\hat Ds^2}=   B^{1}_{112} { S}  { I} \hat \Psi_{I}  + B^{1}_{121}{  S} {  I} \hat \Psi_{S}  +B^{1}_{212} { S} {  I} \hat \Psi_{I}+ B^{1}_{221} { S} {  I} \hat \Psi_{S}  , \eqno{[32.b]}
$$
and
$$
\frac{\hat D^2 \hat \Psi_{I}}{ \hat D s^2} =   B^{2}_{112} { S } { I} \hat \Psi_{I}  + B^{2}_{121} {t S}  { I}\hat \Psi_{S}  +B^{2}_{212} { S}  { I}\hat \Psi_{I}+ B^{2}_{221} { S} { I} \hat \Psi_{S} , \eqno{[32.c]}
$$
i.e.

$$
\frac{\hat D^2\Psi_{S}}{\hat Ds^2} =   0  , \eqno{[32.d]}
$$
and
$$
\frac{\hat D^2 \hat \Psi_{I}}{ \hat D s^2} =  0  . \eqno{[32.e]}$$

Similarly, applying Antonelli's method to define Volterra's equation in ecology [2-7] , we obtain the its corresponding part of the SI-model in Berwald space,
$$
\frac{d^2 S}{dt^2}+ {\hat \Gamma}^{1}_{11} ( \frac{dS}{dt})^2 + 2 {\hat \Gamma}^{1}_{12}( \frac{dS}{dt})( \frac{dI}{dt})+ {\hat \Gamma}^{1}_{22}(
\frac{dI}{dt})^2= \lambda
 ( \frac{dS}{dt} )  \eqno{[33.a]}
$$
 and
$$
\frac{d^2I}{dt^2}+ {\hat \Gamma}^{2}_{11} (\frac{dS}{dt})^2 + 2{\hat \Gamma}^{2}_{12}(\frac{dS}{dt})(\frac{dI}{dt})+ \hat \Gamma^{2}_{22}(\frac{dI}{dt})^2= \lambda (\frac{dI}{dt}), \eqno{[33.b]}
$$

Thus, the coefficients of its affine connection become:
$$
\hat\Gamma^{1}_{11} = \alpha_{1}-\alpha_{2}L,  \eqno{[34.a]}
$$
$$
\hat\Gamma^{1}_{22} = -( \alpha_{1}-\alpha_{2}L ) , \eqno{[34.b]}
$$
$$
\hat\Gamma^{1}_{12} =\hat\Gamma^{1}_{21}= -( \alpha_{2}+\alpha_{1}L )\eqno{[34.c]}
$$
$$
\hat\Gamma^{2}_{12} =\hat\Gamma^{1}_{21}= -( \alpha_{1}- \alpha_{2}L )\eqno{[34.d]}
$$
$$
\hat\Gamma^{2}_{11} = -( \alpha_{2}+\alpha_{1}L )\eqno{[34.e]}
$$
$$
\hat\Gamma^{2}_{22} = -( \alpha_{2}+\alpha_{1}L ) .\eqno{[34.f]}
$$
Thus, the components of path equations are  expressed as follows:

$$
\frac{d S}{dt}+ 2(\beta+ \alpha L) S  I +(\beta L -\alpha)( I)^2 + (\alpha- \beta L)(S)^2= \lambda  S,\eqno{[35.a]}
$$
and
$$
\frac{d I}{dt}+ 2(\alpha- \beta L) S  I -(\beta + \alpha L) S^2+ (\alpha- \beta L)I^2= \lambda  I.\eqno{[35.b]}
$$
 Equations (35.a)and (35.b) are  obtained by taking the variation with respect the corresponding deviation vector  of the following modified Bazanski Lagrangian
$$
L= \bar{g}_{ab} {\dot X}^{a} \frac{\hat D \hat \Psi^{b}}{\hat D t} + \lambda \Psi_{a}{\dot X}^{a} \eqno{[36]}.
$$
Also, after some multiplications, we can find  Their  corresponding deviation equations by taking the variation with respect to $\dot X^c$ to  become:

$$
\frac{\hat D^2\Psi_{S}}{\hat Ds^2}+ \lambda \frac{\hat D  \hat \Psi_{S}}{Dt} =   B^{1}_{112} { S}  { I} \hat \Psi_{I}  + B^{1}_{121}{  S} {  I} \hat \Psi_{S}  +B^{1}_{212} { S} { \dot I} \hat \Psi_{I}+ B^{1}_{221} { S} {  I} \hat \Psi_{S}  ,\eqno{[37.a]}
$$
and
$$
\frac{\hat D^2 \hat \Psi_{I}}{ \hat D s^2}+ \lambda \frac{\hat D  \hat \Psi_{I}}{Dt} =   B^{2}_{112} { S } { I} \hat \Psi_{I}  + B^{2}_{121} { S}  { I}\hat \Psi_{S}  +B^{2}_{212} { S}  {\dot I}\hat \Psi_{I}+ B^{2}_{221} { S} { I} \hat \Psi_{S} \eqno{[37.b]},
$$
i.e.
$$
\frac{\hat D^2\Psi_{S}}{\hat Ds^2}+ \lambda \frac{\hat D  \hat \Psi_{S}}{Dt} =  0  \eqno{[37.c]},
$$
and
$$
\frac{\hat D^2 \hat \Psi_{I}}{ \hat D s^2}+ \lambda \frac{\hat D  \hat \Psi_{I}}{Dt} =  0 \eqno{[37.d]},
$$

\subsection{Antonelli-Finsler Metric Function}
In order to generalize the previous metric in (26) Antonelli has modified Finsler metric to become [7]:
$$
F =e^{\phi}\dot ({\sum}^{n}_{i=1}{(\dot x^{i})^{m})}^{\frac{1}{m}},\eqno{[38]}
$$
where $m \geq 2$ and $n=2$ .
Thus their corresponding  components of its affine connection are given as :
$$
\hat{\Gamma}^{1}_{11}= \sigma_{1} -\frac{1}{9} \beta_{1}(\frac{\dot I}{\dot S})^{\frac{4}{3}} , \eqno{[39.a]}
$$
$$
\hat{\Gamma}^{1}_{21}= \frac{4}{9} \beta_{1}(\frac{\dot I}{\dot S})^{\frac{4}{3}}+ \frac{1}{2}\gamma_{12} ,\eqno{[39.b]}
$$
$$
\hat{\Gamma}^{1}_{22}= \frac{2}{9} \beta_{1}(\frac{\dot S}{\dot I})^{\frac{2}{3}} ,\eqno{[39.c]}
$$
$$
\hat{\Gamma}^{2}_{22}= \sigma_{2} -\frac{1}{9} \beta_{2}(\frac{\dot S}{\dot I})^{\frac{4}{3}} .\eqno{[39.d]}
$$

 Consequently, the  Bazanski method to obtain path and path deviation equations will become in the following way:\\
{[i]} The components of Path Equations:
$$
\frac{d\dot S}{dt}+ \sigma_{1}  S^{2} +\sigma_{2}(\frac{m}{m-1})  S  I + \frac {\sigma_{1}}{m-1} (\frac{ I}{ S})^{m-2} {I}^2 = \lambda_{1}  S \eqno{[40.a]}
$$
and
$$
\frac{d I}{dt}+ \sigma_{2}  I^{2} +\sigma_{1}(\frac{m}{m-1})  S  I + \frac{\sigma_{2}}{m-1}  (\frac{\dot I}{\dot S})^{m-2} { I}^2= \lambda_{2} I.\eqno{[40.a]}
$$

{[ii]} The components of Path deviation Equations:
$$
\frac{\hat D^2\Psi_{S}}{\hat Ds^2}+ \lambda \frac{\hat D  \hat \Psi_{S}}{Dt} =  0  ,\eqno{[41.a]}
$$
and
$$
\frac{\hat D^2 \hat \Psi_{I}}{ \hat D s^2}+ \lambda \frac{\hat D  \hat \Psi_{I}}{Dt} =  0 ,\eqno{[41.b]}
$$
It is well known that  the interaction between $ S \&  I$ can be vitally important when $m > 2$ and the cases of increasing dimensions such as the following section to examine the possibility to geometrize SIR models.
\section{Geomertization of SIR Model}
In a similar way, we can extend our study to examine SIR model using the Bazanski method in each Riemannian and Finslerian to become:
\\
{(i)}For Riemannian Geometry

 $$
 \frac{D S}{Dt}=0  \eqno{[42.a]}
 $$

$$
\frac{D  I}{Dt} =\alpha I,\eqno{[42.b]}
$$
and
$$
\frac{D R}{Dt}= 0 , \eqno{[42.c]}
$$
provided that
$$
S+R+I =N \eqno{[43]}
$$
where $R$ is the number of recovered or dead group.

The above equation can easily be obtained by assuming the following Lagrangian:
$$
L=g_{\mu \nu}U^{\mu} \frac{D \Psi^{\nu}}{Dt} + \alpha
_{(\mu)}\Psi^{\nu}U^{\nu} ,~~~~ a,b=1,2,3 \eqno{[44]} $$
where $ \alpha_{(\mu)}$ is an arbitrary constant. \\
From this perspective, we can develop the SIR model  in its geometric version as follows
$$
\frac{DU^{\mu}}{Ds}= \alpha_{(\mu)} U^{\mu} \eqno{[45]}$$

and their corresponding deviation equations become:
$$
\frac{\hat{D}^2 \hat{\Psi}^{\mu}}{\hat{D}t^2} +\alpha_{(\mu)} \frac{\hat D \hat \Psi^{\mu}}{\hat D t}= R^{\mu}_{\nu \rho \sigma}U^{\nu}U^{\rho}\hat{\Psi}^{\sigma}. \eqno{[46]}
$$

{(ii)} For Finslerian Geometry  (Berwald Type):

In a similar way to  equations [42a-45] , we can easily construct the following equations:
$$
 \frac{\hat DS}{\hat D t} =0, \eqno{[47.a]}
$$

$$
\frac{\hat D I}{\hat Dt} =\alpha I,\eqno{[47.b]}
$$
and
$$
\frac{\hat D R}{\hat D t}= 0.\eqno{[47.c]}
$$

Equations [47 a-c] are obtained by taking the action of the following Largangian:
$$
L=g_{\mu \nu}\hat U^{\mu} \frac{\hat D  \hat \Psi^{\nu}}{\hat Dt} + \alpha
_{(\mu)}\hat \Psi^{\nu}U^{\nu} ,~~~~~~ \mu, \nu=1,2,3. \eqno{[48]} $$
We also can find their corresponding deviation equations to become:
$$
\frac{\hat{D}^2 \hat{\Psi}^{\mu}}{\hat{D}t^2} +\alpha_{(\mu)} \frac{\hat D \hat \Psi^{\mu}}{\hat D t}= B^{\mu}_{\nu \rho \sigma}U^{\nu}U^{\rho}\hat{\Psi}^{\sigma} \eqno{[49]}
$$
For a complete description of this  model will be examined in our future work.

\section{Discussion and Concluding Remarks}
The paper deals with geometrization some epidemic models using their corresponding path equations. Also, we have obtained their corresponding path deviation equations from one single Lagrangian for different types of geometries based under a positive definite metric with constant affine connections and negative curvature. This process is based on a geometrized vision of concept of allometry to examine  epidemic curves as by finding their path equations. The paper has also opened the window to impose non conventional types of geometries in future work to describe more complex models of SI or briefly in  SIR model. One of good results is increasing the dimensions with preserving the concept of a positive metric with constant affine connections and negative curvature will include many other epidemic models such as susceptive-exposed-infective-recovered models (SEIR).

 Finally, this study  can be extended  to obtain such an appropriate prospective vision to visualize how an epidemic behaves on a longer times without many parameters as needed in case of the traditional regression analysis.  This is not the optimal case to get an exact form of epidemic curve, but it is a tend to apply such a concept of geometrization. In addition in our future work we may in need to develop these geometries to include some new parameters affecting the SI-model such as transmission rate [12] to become:
$$
\frac{dS}{dt}=-\alpha S^p I^p \eqno{[50]}
$$ can be studied in future to be geometrized.

Also, for the SIR model can also be extended to include some interactions like  population fertility or immigration to be considered in demographic using partial differential equations of McKendrick- von Forester  with boundary conditions [13].
$$
( \frac{\partial} {\partial t} + \frac{\partial} {\partial a} ) n(a,t) = -\mu n(a,t) + I(a,t) ,~~~~ 0<a<\omega, t>0 \eqno{[5]}
$$
such that
$$
n(0,t) =B(t) = \int^{\omega}_{0} m(a)n(a,t)da ,~~~~~ t\geq0 \\
$$
$$
n(a,t) =n_{0}(a) , 0<a<\omega
$$
where  $\omega$ represents the maximum life span of individuals, $B(t)$ is the number of newborn per unit time $t$ , $m(a)$ is a mortality rates and  $\mu (a)$ is a constant. The evolution of density $n(a,t)$ of individuals aged $a$ and time$t$, will open the window to be included geometrically  as non linear connections in Finslerian approach.\\

\section*{Acknowledgements}
I would like to thank Professors P.L. Antonelli , K. Buchner and R. Shimming for their remarks and comments. I also thank my colleague Dr. E. H. Hassan for encouraging me to continue in this field.
\section*{References}
{[1]}Murray,J.D. (1989){\it{Mathematical Biology}}, biomathematics Texts, Springer-Verlag \\
{[2]} Antonilli  P.L.(1980) Acta Cient. Venezolana 31, 521. \\
{[3]} Antonelli, P.L. (1991) Tensor N.S. 50, 22. \\
{[4]}Voorhees B.H. (1983)  International Journal of Theoretical Physics,  {\bf22},251 \\
{[5]} Antonelli, P.L. and Voorhees B.H. (1983) Bulletin of Mathematical Biology {\bf 45}, 103.\\
{[6]}Antonelli, P.L. and Rutz, S.F. (2009)  Nonlinear analysis Real world applications, {\bf 10 }, 576. \\
{[7]} Antonelli,P.I., Ingarden, R.S. and Matsumoto, M. (1993) {\it{The Theory of Sparys and Finsler Spaces with Applications in Physics and Biology}}, Kluwer Academic Publishers. \\
{[8]} Epstein, J.M. (1997) {\it{ Nonlinear Dynamics,Mathematical Biology and Social Science  }} \\
{[9]} Istas,J.   (2005) {\it{Mathematical Modeling for the Life Sciences}}Springer \\
{[10]} Bazanski, S.L. (1989) J. Math. Phys., {\bf{30}}, 1018. \\
{[11]} Kahil, M.E. (2006), J. Math. Physics {\bf{47}},052501. \\
{[12]} Novozhilov,A. (2008) arXiv:0809.1578 \\
{[13]} Franceschetti,A. and Pugliese,A. (2008) J.Math.Biol. {\bf57},1. \\

\end{document}